\documentclass[bst/sn-basic]{sn-jnl}

\usepackage{graphicx}%
\usepackage{multirow}%
\usepackage{amsmath,amssymb,amsfonts}%
\usepackage{amsthm}%
\usepackage{mathrsfs}%
\usepackage[title]{appendix}%
\usepackage{xcolor}%
\usepackage{textcomp}%
\usepackage{manyfoot}%
\usepackage{booktabs}%
\usepackage{algorithm}%
\usepackage{algorithmicx}%
\usepackage{algpseudocode}%
\usepackage{listings}%
\usepackage{pifont}
\usepackage{natbib}
\usepackage{xcolor}
\usepackage{orcidlink}

\theoremstyle{thmstyleone}%
\theoremstyle{thmstyletwo}%

\theoremstyle{thmstylethree}%

\begin{document}

\title[Article Title]{Benchmarking Histopathology Foundation Models for Ovarian Cancer Bevacizumab Treatment Response Prediction from Whole Slide Images}


\author[1]{\fnm{Mayur} \sur{Mallya} \orcidlink{0000-0002-9432-4262}} \email{mayur.mallya@ubc.ca}

\author[1]{\fnm{Ali} \sur{Khajegili Mirabadi} \orcidlink{0000-0002-0489-6583}} \email{ali.mirabadi@ubc.ca}

\author[2]{\fnm{Hossein} \sur{Farahani} \orcidlink{0000-0002-9503-1875}} \email{h.farahani@ubc.ca}

\author*[2,3,4]{\fnm{Ali} \sur{Bashashati} \orcidlink{0000-0002-4212-7224}} \email{ali.bashashati@ubc.ca}

\affil[1]{\orgdiv{Faculty of Science}, \orgname{University of British Columbia}, \orgaddress{\street{2207 Main Mall}, \city{Vancouver}, \postcode{V6T 1Z4}, \state{British Columbia}, \country{Canada}}}

\affil[2]{\orgdiv{School of Biomedical Engineering}, \orgname{University of British Columbia}, \orgaddress{\street{2222 Health Sciences Mall}, \city{Vancouver}, \postcode{V6T 2B9}, \state{British Columbia}, \country{Canada}}}

\affil[3]{\orgdiv{Department of Pathology and Laboratory Medicine}, \orgname{University of British Columbia}, \orgaddress{\street{2211 Wesbrook Mall}, \city{Vancouver}, \postcode{V6T 1Z7}, \state{British Columbia}, \country{Canada}}}

\affil[4]{\orgdiv{Canada's Michael Smith Genome Sciences Centre}, \orgname{BC Cancer Research Institute}, \orgaddress{\street{570 W 7th Ave}, \city{Vancouver}, \postcode{V5Z 4S6}, \state{British Columbia}, \country{Canada}}}

\abstract{\textbf{Purpose:} Bevacizumab is a widely studied targeted therapeutic drug used in conjunction with standard chemotherapy for the treatment of recurrent ovarian cancer. While its administration has shown to increase the progression-free survival (PFS) in patients with advanced stage ovarian cancer, the lack of identifiable biomarkers for predicting patient response has been a major roadblock in its effective adoption towards personalized medicine.
\textbf{Methods:}  In this work, we leverage the latest histopathology foundation models trained on large-scale whole slide image (WSI) datasets to extract ovarian tumor tissue features for predicting bevacizumab response from WSIs.
\textbf{Results:}  Our extensive experiments across a combination of different histopathology foundation models and multiple instance learning (MIL) strategies demonstrate capability of these large models in predicting bevacizumab response in ovarian cancer patients with the best models achieving an AUC score of 0.86 and an accuracy score of 72.5\%. Furthermore, our survival models are able to stratify high- and low-risk cases with statistical significance (p $<$ 0.05) even among the patients with the aggressive subtype of high-grade serous ovarian carcinoma. 
\textbf{Conclusion:} This work highlights the utility of histopathology foundation models for the task of ovarian bevacizumab response prediction from WSIs. The high-attention regions of the WSIs highlighted by these models not only aid the model explainability but also serve as promising imaging biomarkers for treatment prognosis. 

}




\maketitle

\section{Introduction}\label{sec1}

Ovarian cancer is a highly lethal gynecologic disease and one of the leading causes of cancer-related deaths among women. In 2020, a total of 313,959 new cases of ovarian cancer were recorded globally that resulted in 207,252 new deaths~\cite{huang2022worldwide}. Among these epithelial ovarian carcinoma (EOC) is the most common type accounting for about 90$\%$ of all ovarian malignancies~\cite{torre2018ovarian}. EOC is heterogeneous disease with distinct subtypes each with their own clinical and molecular characteristics that can impact the prognosis and treatment response. High-grade serous carcinoma (HGSC) is the most common and aggressive subtype of EOC with a disproportionate share of fatalities. 

Standard treatment for a newly diagnosed EOC case involves a surgical cytoreduction followed by paclitaxel and platinum-based chemotherapy treatment. However, 70$\%$ of the cases are diagnosed at an advanced stage when the treatment options are not only limited but also ineffective, leading to treatment resistance and tumor recurrence and thereby contributing to poor prognosis and patient outcomes~\cite{burges2011ovarian}. The limited responsiveness to conventional chemotherapy at an advanced stage coupled and the resulting high mortality rate has necessitated the usage of targeted therapeutic agents in the treatment of EOC.

Bevacizumab (clinically known as Avastin) is an extensively studied targeted therapeutic agent used in the treatment of recurrent EOC~\cite{monk2013integrating,garcia2020bevacizumab}. It plays an important role in the inhibition of tumor angiogenesis by neutralizing the vascular endothelial growth factor (VEGF), a key signaling protein responsible for tumor regrowth. The administration of bevacizumab in conjunction with chemotherapy has shown to increase the progression-free survival (PFS) in advanced stage EOC~\cite{burger2011incorporation,perren2011phase}. However, the lack of effective biomarkers for predicting patient responses remains as a challenge for the personalization of bevacizumab therapy to this day. Additionally, given the high costs and potential adverse side-effects, identifying EOC patients with favorable response to bevacizumab therapy becomes an imperative task.

In this work, we use the publicly available ovarian bevacizumab response dataset of histopathological whole slide images (WSI) to predict the treatment effectiveness for the patient~\cite{wang2022histopathological}. Analysis of hematoxylin and eosin (H\&E) stained WSIs is a cost-effective and routine practice in clinical pathology, offering valuable insights into the tumor microenvironment and morphology. The WSIs play an integral role in the clinical diagnosis and management of ovarian cancer. Advancements in deep learning (DL) over the last decade have enabled the computational analysis of WSIs, which are gigapixel resolution and often lack detailed annotations from pathologists. Weakly-supervised learning strategies such as multiple instance learning (MIL) have been prominently applied and have shown great success across a wide variety of WSI analysis tasks such as grading~\cite{su2022attention2majority}, subtyping~\cite{shao2021transmil,lu2021data}, survival analysis~\cite{yao2020whole,liu2024advmil}, etc. By breaking down the WSI into a bag of multiple patches, MIL methods can automatically identify the most discriminative patches and aggregate their features to generate slide-level features which can be used for downstream tasks.

Due to the small-size nature of medical image datasets, including those in histopathology, prior works on ovarian bevacizumab response prediction have either relied on extensive WSI pre-processing strategies for efficient patch selection from the high-resolution WSIs~\cite{wang2022weakly1} or leveraged more informative molecular counterparts for this task~\cite{wang2022weakly2,wang2023ensemble3}. A popular approach for dealing with small-size datasets is using transfer learning strategies where models pre-trained on natural image datasets such as ImageNet are used as feature extractors for histopathology images~\cite{aitazaz2023transfer}. Despite looking significantly different from histopathology images, the ImageNet pre-trained models provide a strong backbone network which can be fine-tuned for task-specific applications. However, the past 2 years have seen the dramatic rise of histopathology foundation models that are trained on massive amounts of tissue data in a self-supervised fashion, thereby providing strong domain-specific feature extractors for histopathology. These models have shown to outperform the models pre-trained on natural images across a wide variety of primary histopathology analysis tasks such as tumor detection, grading, and subtyping across pan-cancer data~\cite{uni,phikon,plip,lunit,ctranspath}. However, to the best of our knowledge, their efficacy on secondary tasks such as treatment response prediction, mutation prediction, etc. from WSIs is not fully explored. 

Our contribution in this work is 3 folds. 1) We provide the first comprehensive benchmark for evaluating the performance of histopathology foundation models across multiple MIL frameworks for the task of bevacizumab treatment response prediction. Our study highlights the superior performance of histopathology foundation models in comparison to the models pre-trained on natural image datasets emphasizing the utility of domain-specific encoders for a relatively unexplored secondary histopathology analysis task of treatment response prediction. 2) Our models are able to identify the patients with favorable response to bevacizumab therapy with an AUC score of 0.86 and an accuracy of 72.5\% when no effective clinical biomarkers exist for predicting the treatment effectiveness for this task. Furthermore, our survival analysis experiments demonstrate statistically significant risk stratification between high- and low- risk cases even among the patients with the highly aggressive HGSC subtype. 
3) Our models identify high-attention tumorous areas in the WSIs providing a promising approach for the identification of prognostic imaging biomarkers for bevacizumab treatment response in ovarian cancer patients.




\section{Materials and Methods}\label{sec_materials_and_methods}

\subsection{Histopathology Foundation Models}

Self-supervised learning has fueled the recent development of foundation models by leveraging vast amounts of unlabeled data which are commonly found in digital pathology. Trained on hundreds of thousands of tissue patches, typically spanning across multiple cancer types, these models are able to learn powerful task-agnostic tissue representations. CTransPath~\cite{ctranspath} and Lunit~\cite{lunit} foundation models trained on the entire TCGA cohort~\cite{weinstein2013cancer} have shown improvements across a variety of primary analysis tasks such as tumor subtyping, mitosis detection, and cell segmentation. Huang \textit{et al.} fine-tuned the CLIP~\cite{clip} model using the histopathology text-image pairs from Twitter to produce PLIP~\cite{plip} which improved the tumor detection and tissue grading and subtyping tasks. UNI~\cite{uni} and Virchow~\cite{vorontsov2023virchow} are trained on some of the largest private data cohorts with 100,000 and 1,000,000 WSIs respectively. Among these, only Phikon~\cite{phikon} and Virchow~\cite{vorontsov2023virchow} explored their performance on a secondary analysis task of mutation prediction while only Phikon explored the task of survival prediction. However, their tests on secondary analysis tasks were limited and didn't produce unanimous conclusions on survival prediction task.

\begin{table}[h]
\caption{Summary of the histopathology foundation models.}
\label{table_fm}
\centering
\begin{tabular}{lcccc}
\toprule
Model & Public & Training data & Cohort size & Model size \\
\cmidrule(lr){1-5}
CTransPath~\cite{ctranspath} & \checkmark & TCGA + PAIP & 32K & 28M\\
Lunit-Dino~\cite{lunit} & \checkmark & TCGA + TULIP & 37K & 22M\\
Phikon~\cite{phikon} & \checkmark & TCGA & 6K & 86M\\
PLIP~\cite{plip} & \checkmark & OpenPath & --  & 86M\\
UNI~\cite{uni} & \checkmark & Mass-100K & 100K  & 307M\\
Virchow~\cite{vorontsov2023virchow} & \text{\sffamily X} & MSKCC & 1.5M  & 632M\\
\bottomrule
\end{tabular}
\footnotetext{In addition to TCGA, the publicly available datasets include PAIP~\cite{kim2021paip} and OpenPath~\cite{plip}.}
\end{table}

\subsection{Multiple Instance Learning}

The weakly-supervised learning nature of MIL models suits perfectly for the analysis of local regions-of-interest in gigapixel WSIs. The MIL-based deep models have seen tremendous success in computational pathology in the recent years~\cite{gadermayr2024multiple}. Ilse \textit{et al.}~\cite{ilse2018attention} proposed the first learnable attention-based aggregation strategy in MIL models (ABMIL) that outperformed the traditional pooling methods. Subsequent works such as CLAM~\cite{lu2021data} and VarMIL~\cite{schirris2022deepsmile} built on top of this method by improving latent representations of the patch and slide features. The onset of self-attention has enabled the integration of transformer-based aggregation strategies in MIL models such as TransMIL~\cite{shao2021transmil}. The consistent success of MIL models have made it a default strategy for the analysis of WSIs and in this work we use the aforementioned MIL methods that have been used extensively in the literature.

\subsection{Dataset}

\begin{figure}[t]
\includegraphics[width=\textwidth]{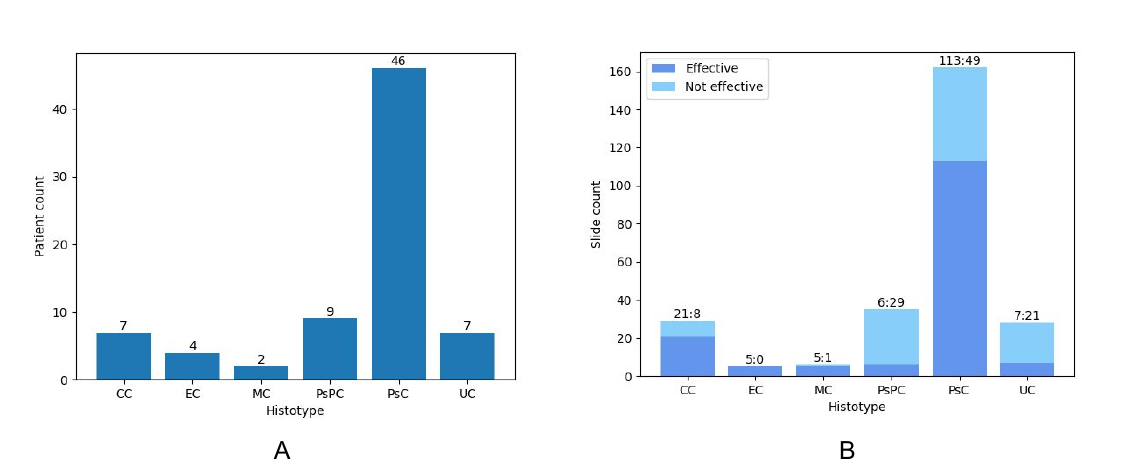}
\caption{a) Patient distribution across the different subtypes in the ovarian bevacizumab response dataset. b) Slide distribution across the different subtypes showing the label distribution for each subtype. The ovarian cancer subtypes include clear cell carcinoma (CC), endometrioid carcinoma (EC), mucinous carcinoma (MC), peritoneal serous papillary carcinoma (PsPC), papillary serous carcinoma (PsC), unclassifed carcinoma (UC). The slide labels include ``Effective'' corresponding to the patients with a favorable response to bevacizumab treatment and ``Not effective'' corresponding the non-responders of the bevacizumab treatment.}
\label{data_distribution}
\end{figure}

Ovarian bevacizumab response dataset~\cite{wang2022histopathological} is a publicly available dataset provided by the Cancer Imaging Archive (TCIA). The dataset consists of 286 hematoxylin and eosin (H\&E) stained whole section WSIs from 78 patients scanned at $20\times$ magnification from the tissue bank of the Tri-Service General Hospital and the National Defense Medical Center, Taipei, Taiwan. The dataset also includes the clinical information of the patients such as the PFS along with the treatment effectiveness of the bevacizumab treatment. The ground-truth binary treatment response was identified based on the pre- and post-treatment CA-125 concentrations, with the bevacizumab treatment being effective for 160 patient slides and ineffective for the remaining 126 slides in the cohort. Fig.~\ref{data_distribution}a. shows the patient distribution across the different ovarian cancer subtypes in the dataset and Fig.~\ref{data_distribution}b. shows the slide-level distribution along with the subtype-wise binary effectiveness label splits.

For training the model, we divide the dataset into 3 folds with training (70\%) and validation (15\%) splits, and we use the same held-out testing split (15\%) to evaluate the models across all folds. Our test set has an equal class distribution with 37 slides each in the effective and ineffective treatment classes (total test set size = 74 slides). The data splits are done at the patient level so that all slides from a patient belong to the same split. Additionally, we conduct experiments exclusively on the serous subtype (includes peritoneal serous papillary carcinoma and papillary serous carcinoma) which is the majority subtype in the dataset as shown in Fig.~\ref{data_distribution}.

\subsection{Problem Formulation}

\begin{figure}[t]
\includegraphics[width=\textwidth]{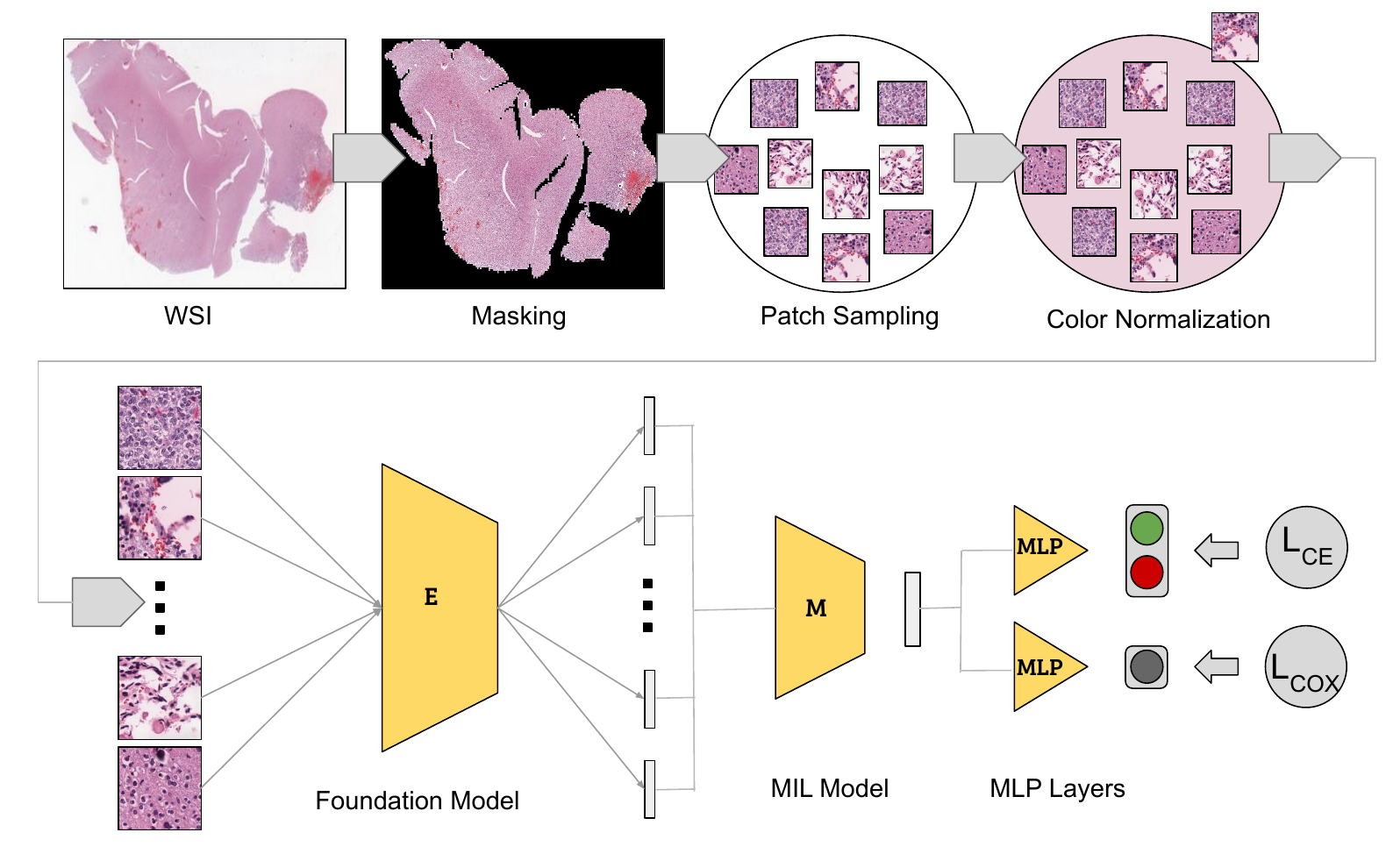}
\caption{Whole slide image (WSI) analysis pipeline depicting the steps involved in the processing of WSIs that includes pre-processing (tissue area masking, patch sampling, color normalization), patch-level feature extraction from foundation models, aggregating patch-level features to produce a slide-level representation using MIL model, and prediction of treatment response. Note that we only train the MIL model along with the MLP layers while training our models with the cross-entropy loss ($L_{CE}$) for a binary classification task of treatment effectiveness prediction or the Cox partial likelihood loss ($L_{COX}$) for the time-to-event regression task of survival prediction.} \label{mil_pipeline}
\end{figure}

We formulate the problem of treatment response prediction from WSIs using two approaches. First is the binary classification approach where the model predicts the treatment effectiveness from WSI whether the treatment was effective ($y=1$) or ineffective ($y=0$). Second is the survival prediction problem where the model uses the WSIs to predict the hazard score of the patient relative to other patients in the cohort. 

We denote the WSI as $W$ and the corresponding binary treatment label, time-to-event, and censor status as $y$, $t$, and $e$ respectively. Our dataset can then be represented as $\{(W_1, y_1, t_1, e_1), (W_2, y_2, t_2, e_2), \cdots (W_n, y_n, t_n, e_n)\}$, where $n$ is the total number of WSIs in the dataset. Each WSI can be treated as a bag of tissue patches denoted by $p$ which can be represented as $W = \{ p_1, p_2 \cdots p_k \}$, where $k$ is the number of patches extracted from each WSI. As part of the pre-processing step, each patch is passed through the color normalization module where a reference patch is used to normalize the stains across different patches. 

Each patch $p_i$ is passed through a pre-trained histopathology foundation model, denoted by $\textbf{E}(\cdot)$ to produce the corresponding feature representation $f_i$,

\begin{equation}
    f_i = \textbf{E}(p_i; \theta^{*}_E)
\end{equation}

\noindent where $\theta_E$ are the parameters of the pre-trained foundation model and the asterisk indicates the parameters are frozen. Each WSI can now be represented as a bag of patch features as, $W = \{f_1, f_2 \cdots f_k\}$. This bag of features is passed to the MIL model denoted by $\textbf{M}(\cdot)$ in order to aggregate the patch-level features of a WSI $W_i$ to form a slide-level representation $s_i$ in a learnable manner. This can be represented as, 

\begin{equation}
    s_i = \textbf{M}(W_i; \theta_M) = \textbf{M}(\{p_{i1}, p_{i2} \cdots p_{ik}\}; \theta_M)
\end{equation}

\noindent where $\theta_M$ denotes the trainable parameters of the MIL model $\textbf{M}(\cdot)$. The slide-level representation $s_i$ is used for predicting the treatment response for the patient corresponding to the WSI $W_i$. The slide-level representation is passed to the MLP layers denoted by $\textbf{MLP}(\cdot)$ to predict the binary treatment response $\hat{y_i}$ as follows,

\begin{equation}
    \hat{y_i} = \textbf{MLP}(s_i; \theta_{MLP})
\end{equation}

\noindent where $\theta_{MLP}$ denotes the trainable parameters of the MLP layers. In the same way, the MLP layers can be used to predict the logarithmic hazard score $h_i$ from the slide-level representation as follows,

\begin{equation}
    h_i = \text{exp}(\textbf{MLP}(s_i; \theta_{MLP}))
\end{equation}

Given the model predictions $\hat{y_i}$ and $h_i$, we use the ground-truth treatment response $y_i$ 
along with the time-to-event $t_i$ and censor status $e_i$ corresponding to WSI $W_i$ to train the model parameters. For training the classification model, we use the binary cross-entropy loss function $L_{CE}(\cdot)$ and for the survival prediction model, we use the cox negative partial log-likelihood loss function $L_{COX}$~\cite{cox1972regression}, which can be calculated as follows,


\begin{equation}
    L_{CE} = - \frac{1}{N} \sum_{i=1}^{N} y_i \cdot \text{log} (\hat{y_i}) + (1-y_i) \cdot \text{log} (1 - \hat{y_i})
    \label{eq_bce_loss}
\end{equation}

\begin{equation}
    L_{COX} = - \sum_{i = 1 | \forall e_i=1}^{N} \left( \text{log}(h_i) - \text{log}\sum_{j \in R(t_i)} h_j \right)
    \label{eq_cox}
\end{equation}

\noindent Eq.~\ref{eq_bce_loss} shows the binary cross entropy loss function where $N$ denotes the number of WSIs in the training set and Eq.~\ref{eq_cox} denotes the cox negative partial log-likelihood function where $R(t_i) = \{j: t_j \geq t_i \}$ denotes the risk set at time $t_i$. In both cases, we minimize the loss functions while updating the parameters of MIL model $\theta_M$ and MLP layers $\theta_{MLP}$ while the parameters of the encoder model $\theta_E$ are frozen.

\subsection{Implementation}

In our experiments, we use a patch size of $224\times224$ at the original $20\times$ magnification and sample $k=300$ patches from each WSI. We use the Macenko stain normalization technique~\cite{macenko2009method,boschman2022utility} to alleviate the staining disparities across different WSIs by performing the normalization across all extracted patches. Further, for training the models, we use the binary cross-entropy loss (Eq.~\ref{eq_bce_loss}) for the classification setting and the cox partial log-likelihood loss (Eq.~\ref{eq_cox}) for the survival prediction setting. For all our experiments, we use Adam optimizer~\cite{kingma2014adam} with a learning rate of $1\times10^{-3}$ and a weight decay of $1\times10^{-2}$. All models are trained for a maximum of $50$ epochs and the model at the best validation performance epoch during the training process is used for inference. To ensure the statistical stability of models, we train all the models with 10 different random seeds across 3 folds and report the 3-fold average metric with the best-performing seed. We use the accuracy and AUC scores as the metrics to evaluate the performance of the models as our test set has an equal class distribution. To evaluate the survival models, we use the commonly used concordance index (c-index) metric~\cite{harrell1996multivariable} in addition to evaluating the risk stratification using the Kaplan-Meier (KM) curves with the log-rank test.
For all our experiments, we use the NVIDIA GeForce RTX 3090 and RTX A6000 GPUs to train the models.

\section{Results and Discussion}\label{sec_results}

In our experiments, we evaluate the performance of the latest histopathology foundation models on a secondary histopathology image analysis task of ovarian cancer bevacizumab response prediction. We compare the performance of the histopathology foundation models with that of the traditionally used encoders pre-trained on natural image datasets. For the histopathology foundation models, we use the openly accessible models namely Phikon~\cite{phikon}, PLIP~\cite{plip}, UNI~\cite{uni}, Lunit-Dino~\cite{lunit}, and CTransPath~\cite{ctranspath}. For the natural image encoders, we use the convolutional backbone models such as ResNet50~\cite{he2016deep}, DenseNet121~\cite{huang2017densely}, and ConvNeXt~\cite{liu2022convnet} along with the transformer backbone models such as ViT~\cite{dosovitskiy2020image} and Swin~\cite{liu2021swin}. Additionally, we also use the KimiaNet~\cite{riasatian2021fine} models which are trained on histopathology images in a supervised fashion unlike the histopathology foundation models that are trained in a self-supervised manner. We evaluate the performance of all encoders across multiple MIL models and average the results to produce a robust benchmark for this task.

\begin{sidewaystable}
\caption{Results of binary classification of bevacizumab treatment effectiveness prediction from histopathology images. The columns represent the MIL models $\textbf{M}(\cdot)$ and the rows represent the pre-trained encoders $\textbf{E}(\cdot)$. The values represent the percentage accuracy scores of prediction averaged across a 3-fold experiment on a held-out test set. Note that encoders in the top half of the table are trained on natural image datasets while those at the bottom are trained on histopathology datasets.}
\label{table_results}
\centering
\begin{tabular}{lcccccc}
\toprule
Encoder             & ABMIL                   & TransMIL                  & VarMIL                    & CLAM-SB                   & CLAM-MB                   & Average           \\
\cmidrule(lr){1-7}
ResNet50              & 50.00 $\pm$ 0.0           & 50.00 $\pm$ 0.0           & 51.35 $\pm$ 1.9           & 50.00 $\pm$ 0.0           & 50.90 $\pm$ 1.2           & 50.45                  \\
DenseNet121           & 50.45 $\pm$ 0.6           & 53.15 $\pm$ 2.7           & 53.15 $\pm$ 4.4           & 57.21 $\pm$ 3.1           & 55.86 $\pm$ 3.5           & 53.96                  \\
ConvNeXt            & 62.16 $\pm$ 3.8           & 55.41 $\pm$ 3.9           & 65.77 $\pm$ 2.7           & 63.96 $\pm$ 5.5           & 65.77 $\pm$ 3.3           & 62.61                  \\   
ViT                 & 63.06 $\pm$ 5.5           & 53.60 $\pm$ 2.7           & 65.32 $\pm$ 2.7           & 67.57 $\pm$ 1.1           & 65.32 $\pm$ 10.0          & 62.97                  \\
Swin (tiny)         & 65.32 $\pm$ 8.4           & 52.70 $\pm$ 5.0           & 65.77 $\pm$ 2.3           & 66.22 $\pm$ 5.0           & 70.72 $\pm$ 4.9           & 64.14                  \\
Swin                & 64.86 $\pm$ 3.3           & 57.21 $\pm$ 8.9           & 64.86 $\pm$ 3.3           & 68.47 $\pm$ 2.7           & 68.02 $\pm$ 3.3           & 64.68                  \\
\cmidrule(lr){1-7}
KimiaNet            & 50.90 $\pm$ 1.2           & 52.70 $\pm$ 1.1           & 62.16 $\pm$ 7.7           & 53.15 $\pm$ 0.6           & 58.56 $\pm$ 6.7           & 55.49                  \\
KimiaNet (OV)       & 57.21 $\pm$ 5.2           & 54.05 $\pm$ 3.9           & 57.66 $\pm$ 6.0           & 54.50 $\pm$ 4.1           & 59.01 $\pm$ 2.3           & 56.48                  \\
\cmidrule(lr){2-7}
Phikon              & 63.96 $\pm$ 7.3           & 67.12 $\pm$ 2.3           & 67.12 $\pm$ 11.2          & 66.67 $\pm$ 3.8           & 68.02 $\pm$ 4.7           & 66.57                  \\
PLIP                & 65.32 $\pm$ 2.3           & 50.00 $\pm$ 0.0           & 66.22 $\pm$ 1.9           & 69.82 $\pm$ 2.7           & 71.17 $\pm$ 6.0           & 64.50                  \\
UNI                 & \textbf{72.07 $\pm$ 3.0}          & 54.95 $\pm$ 8.0           & 69.37 $\pm$ 9.1           & 65.32 $\pm$ 7.2           & 65.77 $\pm$ 3.2           & 65.49 \\  
Lunit-Dino          & 71.17 $\pm$ 3.1 & \textbf{70.72 $\pm$ 4.9}  & 68.47 $\pm$ 0.1           & 69.82 $\pm$ 2.5           & 65.32 $\pm$ 4.5           & 69.10                  \\   
CTransPath          & 69.37 $\pm$ 3.5          & 63.06 $\pm$ 0.1           & \textbf{72.07 $\pm$ 5.2}  & \textbf{72.52 $\pm$ 6.0}  & \textbf{71.62 $\pm$ 3.9}  & \textbf{69.72}                 \\
\bottomrule
\end{tabular}
\footnotetext{KimiaNet (OV) refers to the pre-trained KimiaNet model fine-tuned on an interal ovarian dataset for subtype classification in a supervised fashion.}
\end{sidewaystable}

\begin{figure}[t]
\includegraphics[width=\textwidth]{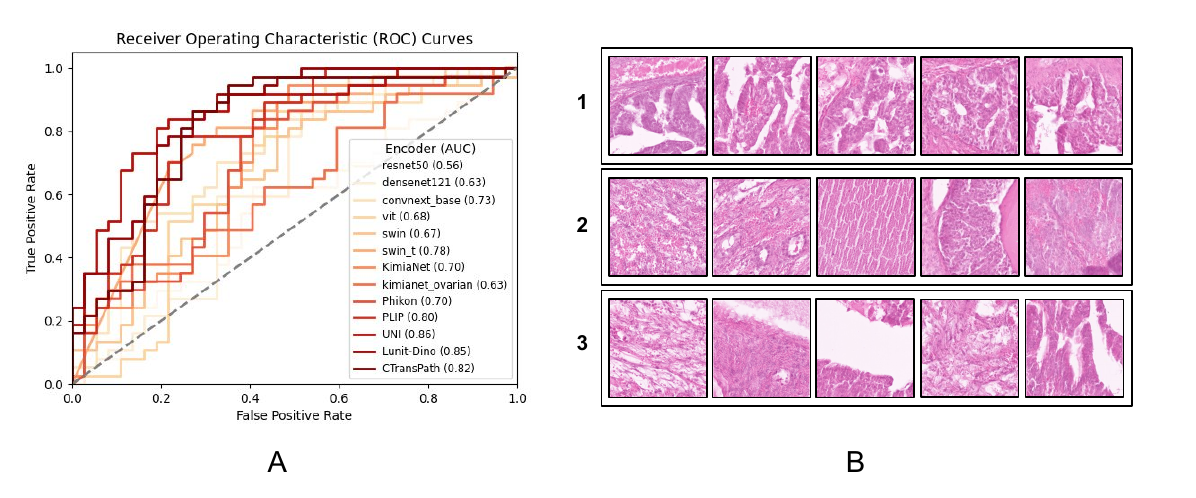}
\caption{a) Receiver Operating Characteristic (ROC) curves showing the treatment response prediction performance along with the AUC score for different encoders. b) Top 5 high-attention patches sorted in the descending order of the attention values (left to right) for the best-performing models from our experiments namely, 1) CTransPath with Clam-SB, 2) CTransPath with VarMIL, 3) Lunit-Dino with ABMIL.} 
\label{insights}
\end{figure}

Table~\ref{table_results} presents the results of binary classification of bevacizumab treatment response prediction on a held-out test set with the values averaged across a 3-fold cross-validation. 
We divide the table into 3 parts -- encoders pre-trained on natural images at the top, encoders pre-trained on histopathology datasets in a supervised manner in the middle, and the histopathology foundation model encoders pre-trained in a self-supervised manner at the bottom of the table. 
The histopathology foundation models achieve the best prediction performance across all MIL frameworks with CTransPath achieving a 72.5\% accuracy with Clam-SB model. Furthermore, the average column showing the average performance of an encoder across MIL models shows that histopathology foundation models achieve superior performance compared to other encoders irrespective of the choice of the MIL framework. Figure~\ref{insights}a. shows the ROC curves for the best model from each encoder along with the corresponding AUC scores and similar to Table~\ref{table_results}, the histopathology foundation models have the highest AUC scores with 4 of the 5 models (except Phikon with an AUC score of 0.7) achieving an AUC score of 0.8 and above.

\begin{figure}
\includegraphics[width=\textwidth]{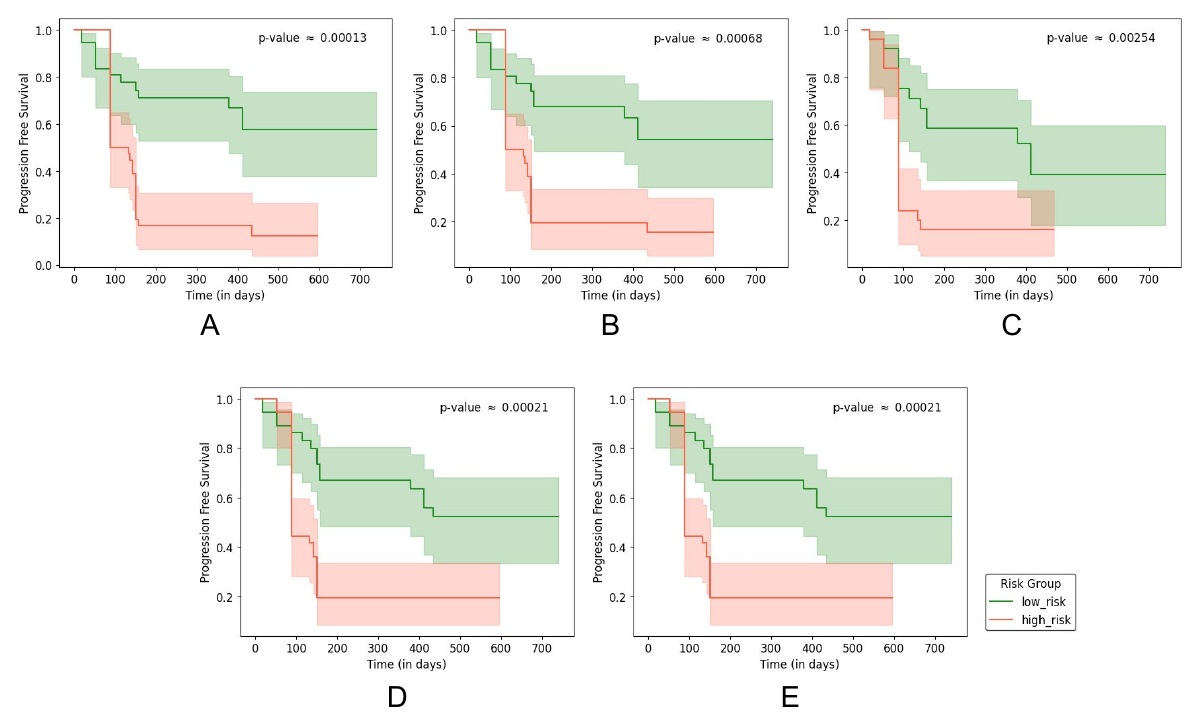}
\caption{Kaplan-Meier plots stratifying the high- and low-risk cases in the test set across the 5 best performing models from Table~\ref{table_survival_results}. a) CTransPath with ABMIL b) CTransPath with VarMIL c) Swin with TransMIL d) UNI with Clam-SB e) UNI with Clam-MB. } 
\label{KM_allsubtypes}
\end{figure}

\begin{figure}
\includegraphics[width=\textwidth]{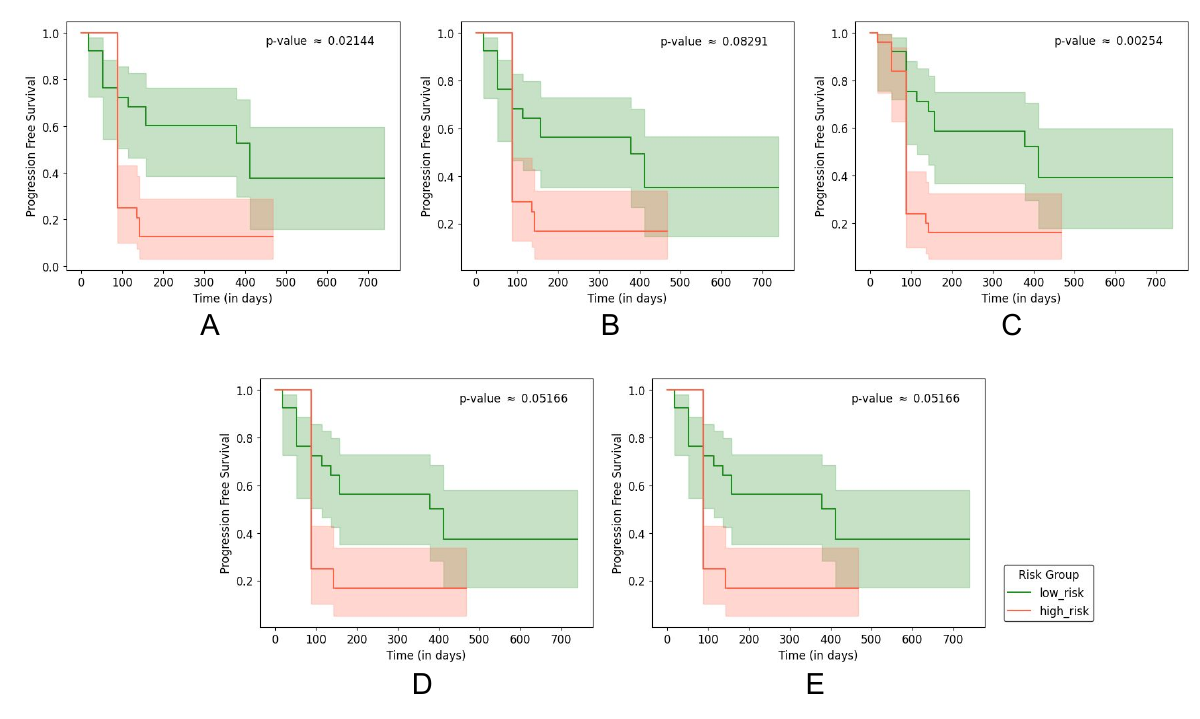}
\caption{Kaplan-Meier plots stratifying the high- and low-risk serous cases in the test set across the 5 best performing models from Table~\ref{table_survival_results}. a) CTransPath with ABMIL b) CTransPath with VarMIL c) Swin with TransMIL d) UNI with Clam-SB e) UNI with Clam-MB.} 
\label{KM_serous_subtypes}
\end{figure}

Table~\ref{table_survival_results} presents the results of bevacizumab treatment survival prediction on a held-out test set averaged across 3-fold cross-validation experiments. Similar to the results of the binary classification in Table~\ref{table_results}, the histopathology foundation models outperform the models pre-trained on natural image datasets on survival prediction task across different MIL models. The histopathology foundation models on average achieve a higher c-index score as compared to the other models with CTransPath and UNI achieving an average c-index of 0.64. Among the natural image encoders, the performance of Swin is on par with the histopathology foundation models with an average c-index of 0.63 while the other natural image encoders are significantly worse compared to the best models. Furthermore, we present the KM survival plots for the best performing models in Table~\ref{table_survival_results} in Figs.~\ref{KM_allsubtypes} and \ref{KM_serous_subtypes} showing the stratification of high- and low-risk cases based on the predicted hazard score. Fig.~\ref{KM_allsubtypes} shows the KM plot corresponding to all cases in the test set while Fig.~\ref{KM_serous_subtypes} shows the KM plot corresponding to only the serous cases of the test set. We use the median PFS value of this cohort as the threshold for differentiating the high- and low-risk cases. In the case of Fig.~\ref{KM_allsubtypes} that includes all subtypes, we observe statistically significant risk stratification on the log-rank test ($p < 0.05$) across all the top-5 best performing models while in the case of Fig.~\ref{KM_serous_subtypes} that includes only the serous subtype, only 2 of the 5 best performing models achieve statistically significant stratification while the remaining 3 models achieve a p-value close to the 0.05 cut-off.

\begin{sidewaystable}
\caption{Results of survival prediction from histopathology images. The columns represent the MIL models $\textbf{M}(\cdot)$ and the rows represent the pre-trained encoders $\textbf{E}(\cdot)$. The values represent the c-index scores of averaged across a 3-fold experiment on a held-out test set.}
\label{table_survival_results}
\centering
\begin{tabular}{lcccccc}
\toprule
Encoder             & ABMIL                       & TransMIL                  & VarMIL                    & CLAM-SB                   & CLAM-MB                   & Average                \\
\cmidrule(lr){1-7}
ResNet50            & 0.59 $\pm$ 0.01               & 0.56 $\pm$ 0.02           & 0.57 $\pm$ 0.02           & 0.59 $\pm$ 0.01           & 0.58 $\pm$ 0.01           & 0.58                  \\
DenseNet121         & 0.54 $\pm$ 0.04               & 0.55 $\pm$ 0.02           & 0.57 $\pm$ 0.03           & 0.57 $\pm$ 0.01           & 0.57 $\pm$ 0.01           & 0.56                  \\
ViT                 & 0.55 $\pm$ 0.02               & 0.60 $\pm$ 0.02           & 0.54 $\pm$ 0.00           & 0.59 $\pm$ 0.03           & 0.59 $\pm$ 0.03           & 0.57                  \\   
Swin                & 0.62 $\pm$ 0.03               & \textbf{0.67 $\pm$ 0.02}  & 0.59 $\pm$ 0.02           & 0.63 $\pm$ 0.01           & 0.63 $\pm$ 0.01           & 0.63                  \\
\cmidrule(lr){1-7}
KimiaNet (OV)       & 0.53 $\pm$ 0.01               & 0.58 $\pm$ 0.01           & 0.56 $\pm$ 0.02           & 0.49 $\pm$ 0.04           & 0.49 $\pm$ 0.04           & 0.53                  \\
\cmidrule(lr){2-7}
Phikon              & 0.59 $\pm$ 0.01               & 0.60 $\pm$ 0.02           & 0.59 $\pm$ 0.02           & 0.60 $\pm$ 0.03           & 0.60 $\pm$ 0.03           & 0.60                  \\
PLIP                & 0.62 $\pm$ 0.02               & 0.63 $\pm$ 0.02           & 0.63 $\pm$ 0.02           & 0.63 $\pm$ 0.01           & 0.63 $\pm$ 0.01           & 0.63                  \\
Lunit-Dino          & 0.63 $\pm$ 0.00               & 0.62 $\pm$ 0.01           & 0.60 $\pm$ 0.01           & 0.59 $\pm$ 0.01           & 0.59 $\pm$ 0.01           & 0.60                  \\   
CTransPath          & \textbf{0.66 $\pm$ 0.03}      & 0.62 $\pm$ 0.02           & \textbf{0.65 $\pm$ 0.03}  & 0.63 $\pm$ 0.01           & 0.63 $\pm$ 0.01           & 0.64                  \\
UNI                 & 0.60 $\pm$ 0.02               & 0.65 $\pm$ 0.03           & 0.63 $\pm$ 0.02           & \textbf{0.67 $\pm$ 0.03}  & \textbf{0.67 $\pm$ 0.03}  & \textbf{0.64}         \\  
\bottomrule
\end{tabular}
\footnotetext{We do not include ConvNext, Swin (tiny), and KimiaNet due to the lack of model convergence due to the \textit{NaN} values in the loss function.}
\end{sidewaystable}

Interestingly, both KimiaNet and KimiaNet-OV (KimiaNet fine-tuned on an internal ovarian dataset for subtype classification in supervised fashion) features fail to perform well on this task showing that supervised pre-training doesn't generalize for tasks outside of its expertise while emphasizing the need for task-agnostic self-supervised histopathology pre-training. Moreover, the KimiaNet models use a convolutional backbone similar to that of ResNet50, DenseNet121, and ConvNeXt while all other encoders in Table~\ref{table_results} use a transformer-based architecture, and the performance disparity due to the choice of model architecture underscores the advantage of transformer-based architecture over convolutional backbones for histopathology analysis task.

Model explainability is an important aspect of medical image analysis and the MIL models in our study use the attention mechanism to rank the patches in the order of their informativeness to the task. Figure~\ref{insights}b. shows the top-5 high attention tumor patches from the 3 best performing models from our experiments in Table~\ref{table_results}. We observe that despite the similar performance, these models assigned different attention to the patches without any common patches among the top-5 high-attention patches across these models. Nevertheless, this approach provides a promising direction for the identification of novel imaging biomarkers for ovarian cancer bevacizumab therapy and we encourage the future works to validate this approach on other ovarian cohorts to confirm the findings. Exploring biological insights from the high-attention patches from different models can also be a valuable research direction for future works.

\section{Conclusion}

Lack of effective biomarkers for predicting the treatment response of ovarian bevacizumab therapy has been a long-standing hurdle in the personalization of its treatment. Additionally, the high costs and potential toxicity make it imperative to identify patient response to ovarian bevacizumab treatment. 
In this work, we perform the first comprehensive benchmarking study evaluating the performance of newly developed histopathology foundation models (along with traditionally used models trained on natural images) on a relatively unexplored secondary histopathology analysis task of ovarian bevacizumab treatment response prediction. 
In addition to achieving a prediction AUC of 0.86 and an accuracy of 72.5\%, our models significantly stratify the low-risk cases from the high-risk ones. Furthermore, the models can identify informative tumor regions in the WSIs corresponding to the treatment response prediction thereby serving as a promising direction for the identification of prognostic imaging biomarkers for bevacizumab treatment of ovarian cancer.

\backmatter





\bmhead{Acknowledgements}
The authors acknowledge the Canada's Michael Smith Genome Sciences Centre for hosting the computational clusters used in this research.


\section*{Declarations}


\bmhead{Funding}

This research was funded by the Natural Sciences and Engineering Research Council of Canada (NSERC), Canadian Institutes of Health Research (CIHR), Health Research BC, and the BC Cancer Foundation.

\bmhead{Competing interests}

The authors declare no competing interests.

\bmhead{Ethics approval and consent to participate}

This research did not require ethics approval and consent to participate as the dataset used in this study is publicly available.

\bmhead{Consent for publication}

Not applicable.

\bmhead{Data availability}

The ovarian bevacizumab response prediction dataset used in this study is publicly available and can be downloaded from the cancer imaging archive (TCIA) database at \url{https://www.cancerimagingarchive.net/collection/ovarian-bevacizumab-response/}


\bmhead{Author contribution}

The project was conceptualized and administered by HF and AB. AKM developed the codebase for running the experiments. MM ran the experiments and wrote the manuscript. All authors reviewed 
and approved the submitted version.

\bibliographystyle{bst/sn-basic}
\bibliography{sn-article}

\end{document}